# High intrinsic lattice thermal conductivity in monolayer MoSi$_2$N$_4$

Jihai Yu[1], Jian Zhou[2*], Xiangang Wan[1,3], and Qingfang Li[4*]

[1] *National Laboratory of Solid State Microstructures and School of Physics, Nanjing University, Nanjing 210093, China*

[2] *National Laboratory of Solid State Microstructures and Department of Materials Science and Engineering, Nanjing University, Nanjing 210093, China*

[3] *Collaborative Innovation Center of Advanced Microstructures, Nanjing University, Nanjing 210093, China*

[4] *Department of Physics, Nanjing University of Information Science & Technology, Nanjing 210044, China*

* Email addresses: zhoujian@nju.edu.cn, qingfangli@nuist.edu.cn

**Abstract:** Very recently, a novel two-dimension (2D) MXene, MoSi$_2$N$_4$, was successfully synthesized with excellent ambient stability, high carrier mobility, and moderate band gap (Science **369**, 670, 2020). In this work, the intrinsic lattice thermal conductivity of monolayer MoSi$_2$N$_4$ is predicted by solving the phonon Boltzmann transport equation based on the first-principles calculations. Despite the heavy atomic mass of Mo and complex crystal structure, the monolayer MoSi$_2$N$_4$ unexpectedly exhibits a quite high lattice thermal conductivity over a wide temperature range between 300 to 800 K. At 300 K, its in-plane lattice thermal conductivity is 224 Wm$^{-1}$K$^{-1}$. The detailed analysis indicates that the large group velocities and small anharmonicity are the main reasons for its high lattice thermal conductivity. We also calculate the lattice thermal conductivity of monolayer WSi$_2$N$_4$, which is only a little smaller than that of MoSi$_2$N$_4$. Our findings suggest that monolayer MoSi$_2$N$_4$ and WSi$_2$N$_4$ are potential 2D materials for thermal transport in future nano-electronic devices.

## I. INTRODUCTION

Since the successful exfoliation of monolayer graphene [1], there have been extensive efforts to find novel two-dimensional (2D) materials due to their unusual mechanical, thermal, optoelectric, piezoelectric, and thermoelectrical properties[2-10]. Owing to these unique properties, 2D materials have become promising candidates for optoelectronics[11], field-effect transistors[12], and energy applications[13,14]. Many classes of monolayer 2D materials have been fabricated, such as transition metal dichalcogenides[2,15], h-BN[16], phosphorene [17], borophene[18], and silicene[19]. It may have difficulties if these monolayer materials are directly applied in integrated nanoelectronic devices because of their restricted properties. For example, graphene and borophene have no band gap[20, 21] while monolayer BN has an excessively wide band gap[22]. Monolayer phosphorene and silicene are unstable upon exposure[19, 23]. The carrier mobility in monolayer $MoS_2$ is quite low[24]. Thus, the discovery of a desirable monolayer 2D material with a moderate band gap and high carrier mobility remains a primary research goal in materials science and physics.

Recently, a high-quality 2D MXene, $MoSi_2N_4$, was successfully synthesized with excellent ambient stability, moderate band gap, and considerable carrier mobility [25]. Experimental results [25] demonstrated that $MoSi_2N_4$ has a band gap of about 1.94 eV. Furthermore, large hole and electron mobilities of monolayer $MoSi_2N_4$ are predicted to be about 1200 and 270 $cm^2V^{-1}s^{-1}$[25], which are 4 to 6 times higher than those of $MoS_2$ monolayer [26]. Since the synthesis of $MoSi_2N_4$, intensive research efforts have been devoted to unearthing its novel properties[27-32]. First principles calculations revealed that the large thermopower of monolayer $MoSi_2N_4$ can be obtained in a range of chemical potential from 0 to 1 eV [27]. Cao *et al* [28] investigated the electrical contact of monolayer $MoSi_2N_4$ and an ultralow Schottky barrier height was observed in $MoSi_2N_4$/$NbS_2$ contact [28], which is beneficial for the nano-electronic device applications. The theoretical calculations demonstrated that the piezoelectricity MoSi$_2$N$_4$ enables actuating new electronic components of nanoscale devices [32].

Thermal transport is an important property for materials in many applications including thermal barrier coating[33], heat management[34], and thermoelectric energy conversion[35]. Especially the phonon transport is an essential part of designing all power-dissipating devices [36, 37]. 2D materials are ideal platforms to investigate fundamental carrier transport and provide new directions for thermal management and energy control. Phonon transport phenomena are related to various intriguing applications based on 2D materials[34]. In recent years, the lattice thermal conductivities of 2D materials have attracted considerable interest [38-50]. For these reasons, the study of the thermal transport property of monolayer MoSi$_2$N$_4$ is urgently called to speed up its application.

In this paper, we have systematically investigated the intrinsic thermal transport properties of monolayer MoSi$_2$N$_4$ by iteratively solving the Boltzmann transport equation. It is found that MoSi$_2$N$_4$ unexpectedly exhibits a quite high lattice thermal conductivity despite its great average atomic mass and complex crystal structure. To further explain the mechanism of the high thermal conductivities, we also discuss the phonon lifetimes, group velocities, and Grüneisen parameters of the monolayer MoSi$_2$N$_4$.

## II. THEORETICAL METHODS

The crystal structure of monolayer MoSi$_2$N$_4$ is fully optimized by the Vienna ab initio simulation package（VASP）[51,52] based on the density functional theory (DFT). The projected augmented wave (PAW) method [53,54] and generalized gradient approximation with the Perdew-Burke-Ernzerh of exchange-correlation functional [55] are used. The plane-wave cutoff energy of 520 eV is used with a 12 × 12 × 1 k-mesh. Both the lattice constants and internal atomic positions are allowed to relax until the maximal residual Hellmann-Feynman forces are less than 0.0001 eV/Å. To avoid interactions with other neighboring layers, a vacuum space of 15 Å is taken.

After optimizing the crystal structure, we further perform the calculations of the second- and third-order interatomic force constants (IFCs) with the finite displacement method. The second-order IFCs in the harmonic approximation and the phonon dispersions of monolayer MoSi$_2$N$_4$ are calculated by using the PHONOPY code[56]. And the third-order IFCs and their lattice thermal conductivities are obtained based on the PHONO3PY code [57], which solves the phonon Boltzmann transport equation by using the iterative self-consistent algorithm. The lattice thermal conductivity is defined as [57]：

$$\kappa = \frac{1}{NV}\sum_\lambda C_\lambda \boldsymbol{v}_\lambda \otimes \boldsymbol{v}_\lambda \tau_\lambda, \tag{1}$$

where N and V are the number of unit cells in the crystal, the volume of a unit cell. $\boldsymbol{v}_\lambda$, and $\tau_\lambda$ are the group velocity and lifetime of the phonon mode $\lambda$, respectively. The method has already been widely used in the prediction of thermal conductivities for 2D materials [39,41,42,49-51]. A 4 × 4 × 1 supercell (112 atoms) is used to calculate the second- and third-order IFCS in monolayer MoSi$_2$N$_4$ with a cutoff distance of 5.0 Å. And a q-mesh of 30 × 30 × 1 is taken for the calculation of lattice thermal conductivities.

## III. RESULTS AND DISCUSSION

### A. Crystal structure and phonon dispersions

The structure of monolayer MoSi$_2$N$_4$ has seven atoms per unit cell. As shown in Fig. 1, the crystal exhibits a sandwiched structure, where the 2H $MoS_2$-type $MoN_2$ layer is sandwiched between two slightly buckled SiN layers. Monolayer $MoSi_2N_4$ holds a mirror paralleling to the horizontal plane, inversion asymmetry, and $C_3$ rotation symmetry. The lattice constants obtained in our calculations are a=b=2.911 Å, which are a little smaller than those of monolayer $MoS_2$ (3.16 Å) [9]. The distances between Si and its adjacent N (N1 and N2) are 1.748 and 1.755 Å, respectively. The distance of Mo-Si is 2.093 Å. The thicknesses ($L_d$) of the vertical $MoSi_2N_4$ plane is 7.00 Å. These results are in good agreement with the previous reports [25, 27]. We also check the electronic properties by calculating its electron band structure, which is given in Fig. S1. It is found that monolayer $MoSi_2N_4$ exhibits an indirect band gap of 1.77 eV with the valence band maximum (VBM) and conduction band minimum (CBM) located at Γ and K points, respectively, which is also consistent with the previous studies[25, 27, 29, 30]. The band gap could be improved to be 2.30 eV based on the HSE functional calculations[25]. The wide band gap implies that the lattice thermal conductivity is dominant in the total thermal conductivity of monolayer $MoSi_2N_4$. It can be seen from the partial density of states in Fig. S1 that the VBM and CBM are mostly dominated by Mo's d orbitals, with a small contribution from N's $p_z$ states. There is a strong hybridization between Mo's d (Si's p) and N's p orbitals from about -10.0 to -2.0 eV and 2.0 to 6.0 eV. The result is consistent with the charge analysis, indicating about 2.2 e (1.5 e) transferring from Si (Mo) atom to its adjacent N atom, respectively [27]. The strong hybridization and large charge transfer suggest that there are strong interactions between Mo (Si) and N atoms.

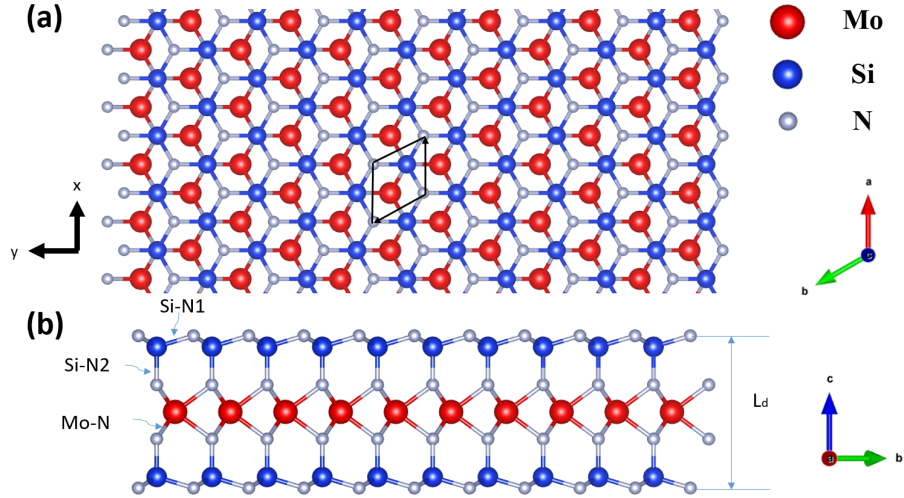

FIG. 1. (a) Top and (b) side views of the crystal structure of monolayer $MoSi_2N_4$. The primitive unit cell is indicated by a black hexagonal in (a). The red, blue, and brown balls represent Mo, Si, and N atoms, respectively.

The phonon dispersions play a significant role in the precise calculation of phonon transport properties. Based on our optimized structures, the phonon dispersions and density of states (DOS) of $MoSi_2N_4$ are calculated by using the PHONOPY code, displayed in Fig. 2. There are no imaginary frequencies in both phonon dispersions and DOS, which means that the structure of monolayer $MoSi_2N_4$ is dynamically stable. There are 21 phonon branches, including 3 acoustic phonon branches and 18 optical phonon ones. The highest phonon frequency is about 31.0 THz (i.e. about 1035 $cm^{-1}$), while the highest frequency of acoustic branches is about 10 THz, which agrees well with the results in Refs. [25, 27]. There are two band gaps of optical branches in $MoSi_2N_4$, which lie in 14.4 to 15.7 THz and 22.0 to 23.7 THz, respectively.

The phonon DOS in Fig.2 (b) indicates that the high-frequency (above 15 THz) phonon modes are mainly attributed to the vibrations of N ions, while some contribution from the Si ions is also found. The vibrational frequencies of Mo ions are mostly below 15 THz due to the greater atomic mass of Mo atom but the Si and N atoms also have a large contribution in this region. The contribution of Mo, Si, and N to the phonon spectrum below 15 THz is 29.4%, 39.9%, and 30.7%, respectively, shown that the SiN layer contributes more to the low-frequency phonon spectrum.

The sound velocities of monolayer $MoSi_2N_4$ are also calculated based on the phonon dispersion, which is given in Table I. The average sound velocity can be determined by the formula $3/v_i^3 = 1/v_{i,LA}^3 + 1/v_{i,TA}^3 + 1/v_{i,ZA}^3$, where $i$ represents the x or y axis. Along the x axis, the sound velocities in the long-wavelength limit are 10.7, 6.1, and 2.0 km/s for the LA, TA, and ZA phonons respectively. These values are higher than those of silicene (8.8, 5.4, and 0.63 km/s for the LA, TA, and ZA phonons) [58] and $MoS_2$ (6.6 and 4.3 for the LA and TA phonons) [58] but much smaller than those of graphene (19.9 and 12.9 km/s for the LA and TA phonons) [59]. The sound velocities of monolayer $MoSi_2N_4$ along the y axis are almost identical to that in the x axis. The large sound velocities are one of the main reasons for the high lattice thermal conductivities of monolayer $MoSi_2N_4$ as we will show later.

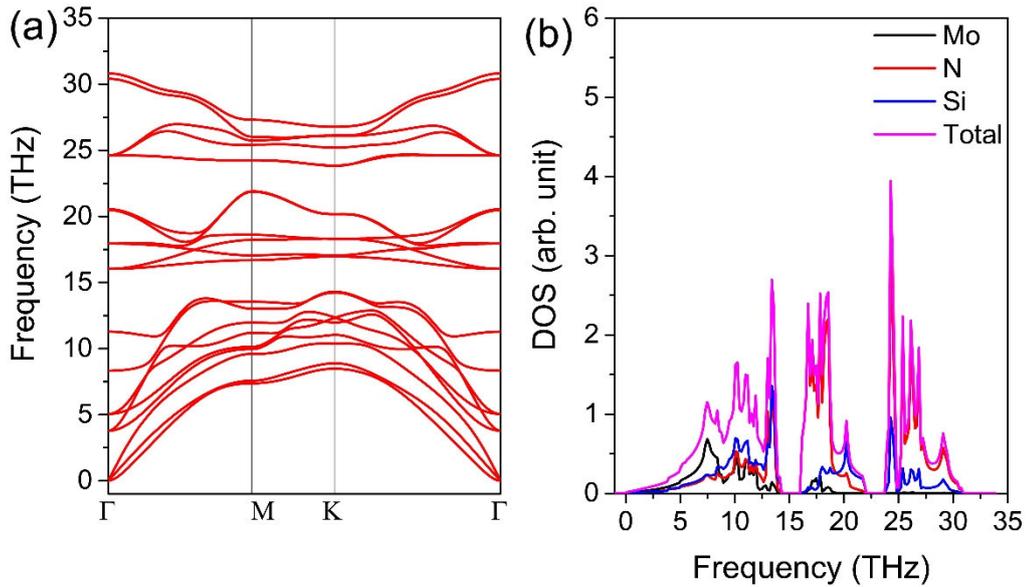

FIG. 2. (a) Phonon dispersions and (b) DOS of monolayer $MoSi_2N_4$.

Table I. Calculated sound velocities of monolayer $MoSi_2N_4$ along x and y axes in the unit of km/s.

|        | LA   | TA  | ZA  | average |
|--------|------|-----|-----|---------|
| x axis | 10.7 | 6.1 | 2.0 | 2.9     |
| y axis | 10.8 | 6.1 | 2.2 | 3.1     |

**B. Lattice thermal conductivity**

We then calculate the temperature-dependent lattice thermal conductivity of monolayer $MoSi_2N_4$, as depicted in Fig. 3. The intrinsic in-plane lattice thermal conductivity decrease with the increase of temperature, which could be explained by the Umklapp scattering mechanism [60]. At 300 K, the lattice thermal conductivity of $MoSi_2N_4$ are 224 $Wm^{-1}K^{-1}$, which is much higher than those of the other well-known 2D semiconductors, such as black phosphorene (30.15 $Wm^{-1}K^{-1}$ (zigzag), 13.65 $Wm^{-1}K^{-1}$ (armchair)) [47], monolayer $2H-MoTe_2$ (42.2 $Wm^{-1}K^{-1}$) [48], $MoS_2$ (83 $Wm^{-1}K^{-1}$ [49] or 23.2 $Wm^{-1}K^{-1}$ [61]) and blue phosphorene (106.6 $Wm^{-1}K^{-1}$) [50], while much lower than that of monolayer hexagonal BN, BP, BAs [62], $C_3N$ [63], and graphene [64] with low average atomic mass. It is noted that the thermal conductivity of $MoSi_2N_4$ is even much higher than those of widely used electronic materials such as Si (142 $Wm^{-1}K^{-1}$) [65]. Hence, the satisfactory lattice thermal conductivity of $MoSi_2N_4$ could guarantee heat removal in the corresponding nano-electronic devices.

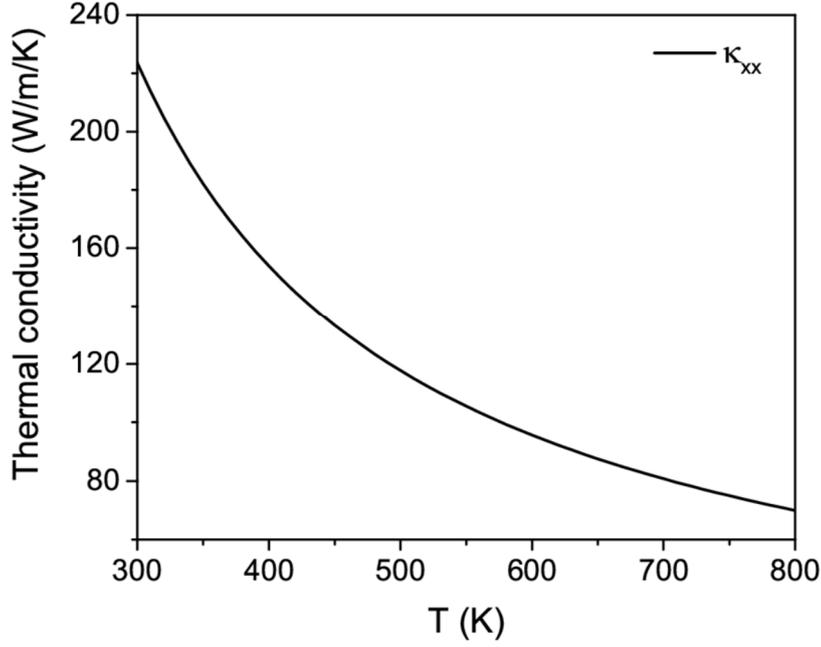

FIG. 3. Calculated in-plane lattice thermal conductivity ($\kappa_{xx}$) of monolayer $MoSi_2N_4$ from 300 to 800 K in the unit of $Wm^{-1}K^{-1}$.

To deeply understand the lattice thermal conductivity of monolayer $MoSi_2N_4$, we then further calculate the cumulative lattice thermal conductivity of $MoSi_2N_4$ at 300 K, given in Fig.4 (a). The cumulative lattice thermal conductivity first increases with the increase of MFP, and then gradually saturates when the phonon mean free path (MFP) is equal to or larger than 1000 nm, which is much longer than those of black phosphorene (66/83 nm) [47], but shorter than graphene [64]. The phonon MFP in $MoSi_2N_4$ is much longer than those in the other 2D materials, leading to a much higher thermal conductivity. The representative MFP (rMFP) of materials is useful for studying the size effects on the diffusive or ballistic phonon transport. The rMFP means the phonons whose MFP is smaller than their rMFP contribute to half of the total lattice thermal conductivity. The rMFP of $MoSi_2N_4$ is 156.3 nm, and the values are about ten times that of phosphorene (17/15 nm [47]).

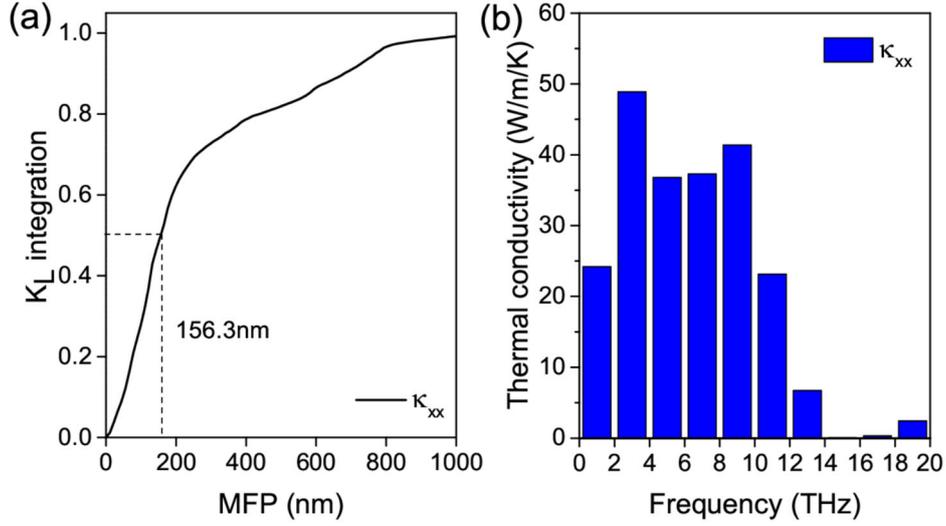

FIG. 4. (a) Normalized directional cumulative lattice thermal conductivity ($K_L$) with respect to the phonon mean free path (MFP) and (b) frequency-dependent in-plane lattice thermal conductivity of monolayer $MoSi_2N_4$ at 300 K. In (b), the phonons above 20 THz contribute little to the thermal conductivity, which is not shown here.

We also calculated the frequency-dependent lattice thermal conductivity of monolayer $MoSi_2N_4$ at 300 K, which is presented in Fig. 4(b). The width of each column in the figure is 2.0 THz. The summation of all columns represents the total thermal conductivity. It is found that the phonon below 15 THz contribute most of (96%) the lattice thermal conductivity in monolayer $MoSi_2N_4$. Furthermore, we also analyzed the contribution of acoustic and optical phonons. It is found that the acoustic phonons contribute about 55% and the optical ones contribute about 45% of the thermal conductivity in both directions. The large contribution of the optical phonons is due to the large number of low-frequency optical modes of the complex structure of monolayer $MoSi_2N_4$.

**C. Phonon group velocities, lifetimes, and Grüneisen parameters**

To understand the underlying mechanism of the high intrinsic lattice thermal conductivity in MoSi$_2$N$_4$, we further analyze its phonon group velocities, lifetimes, and Grüneisen parameters. The squares of the phonon group velocities are plotted in Fig. 5 (a). The large values of squares of the group velocities almost lie below about 15 THz which could reach more than 100 km$^2$/s$^2$ which are much larger than those of monolayer MoS$_2$ [61]. Since the lattice thermal conductivity is proportional to the squares of group velocities, therefore the large group velocities in Fig. 5(a) as well as the large sound velocities in Table I are the important reasons for its high thermal conductivity.

The frequency-dependent phonon lifetimes of MoSi$_2$N$_4$ are calculated by using the PHONO3PY code from the third-order anharmonic IFCs, as displayed in Fig. 5 (b). The phonon lifetimes at the low frequency (acoustic phonon modes) are much longer than those of high-frequency optical modes. Most of the phonon lifetimes in MoSi$_2$N$_4$ are lie in the range from 0 to 50 ps. The lifetimes are larger than those of monolayer MoS$_2$ [61], and even higher than those of penta-graphene (PG) and hydrogenated PG (HPG), while the lattice thermal conductivities of PG (350 Wm$^{-1}$K$^{-1}$) [66] and HPG (616 Wm$^{-1}$K$^{-1}$) [66] are much larger than that of MoSi$_2$N$_4$. The results imply that the long phonon lifetimes contribute significantly to the large thermal conductivity of monolayer MoSi$_2$N$_4$. Finally, we give the mode-dependent Grüneisen parameters ($\gamma$) of monolayer MoSi$_2$N$_4$, which can provide crucial information on the anharmonic interactions of phonons. The larger $\gamma$ implies stronger anharmonicity, which leads to low lattice thermal conductivity. Fig. 5 (c) indicates that Grüneisen parameters of MoSi$_2$N$_4$ are dominantly located in the range from -1.5 to 1.5. The range is smaller than those of PG, HPG, and graphene (-8~2) [66] which have ultra-high lattice thermal conductivities. The long lifetimes and small Grüneisen parameters indicate the small anharmonicity in monolayer MoSi$_2$N$_4$, which is another important reason for its high lattice thermal conductivity. This weak anharmonicity is attributed to the strong Mo-N and Si-N atomic interactions since Young's modulus of MoS$_2$N$_4$ (479 GPa) is much higher than that of MoS$_2$ (270 GPa). [25]

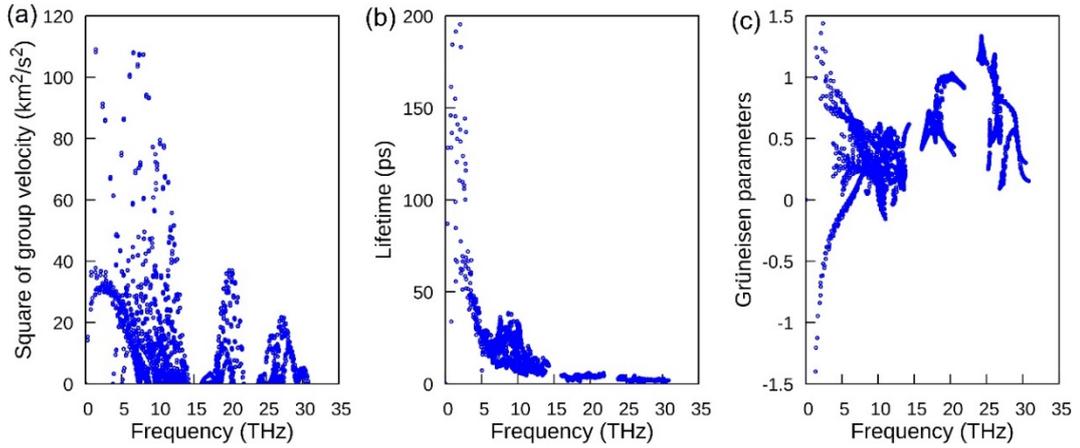

FIG. 5. (a) Square of the group velocities, (b) phonon lifetimes, and (c) frequency-dependent mode Grüneisen parameters of monolayer MoSi$_2$N$_4$ at 300 K.

**D. Comparative study with monolayer WSi$_2$N$_4$**

Finally, we compare MoSi$_2$N$_4$ with another 2D MXene (WSi$_2$N$_4$), which has also been successfully synthesized in the recent experiment[25]. Compared with monolayer MoSi$_2$N$_4$, WSi$_2$N$_4$ has the same crystal structure and similar band characteristics[25], but a wider band gap[25] and higher atomic density. To understand the difference of phonon transport properties between WSi$_2$N$_4$ and MoSi$_2$N$_4$, we further calculate the lattice thermal conductivity of monolayer WSi$_2$N$_4$, as given in Fig. S2. Monolayer WSi$_2$N$_4$ also exhibits high lattice thermal conductivity. At 300 K, its lattice thermal conductivity is 219 Wm$^{-1}$K$^{-1}$. We notice that although the atomic mass of W is much larger than that of Mo, the lattice thermal conductivity of WSi$_2$N$_4$ is only slightly lower than that of MoSi$_2$N$_4$. The similar thermal conductivity results from their similar phonon spectra, the square of the group velocities, phonon lifetimes, and Gruneisen parameter as shown in Fig. S3.

**E. Discussion**

In 1972, G. A. Slack investigated many nonmetallic crystals with high thermal conductivity (>100 Wm$^{-1}$K$^{-1}$ at 300 K) and found four characteristics of them: (1) low average atomic mass, (2) simple crystal structure, (3) strong interatomic bonding, and (4) low anharmonicity [67]. In our case, MoSi$_2$N$_4$ and WSi$_2$N$_4$ have quite complex crystal structures (seven atoms in a unit cell) and great average atomic masses. However, they unexpectedly exhibit quite high lattice thermal conductivities (> 200 Wm$^{-1}$K$^{-1}$ at 300 K). We think that the high thermal conductivities of MoSi$_2$N$_4$ and WSi$_2$N$_4$ are possibly due to their particular sandwiched structure, in which the heavy W or Mo layer is the inner layer while the light Si-N layers are the outer layers. The inner and outer layers are in parallel connection and therefore the heat could still transfer fast in the Si-N layers despite the heavy Mo or W layers. This is possibly also the reason why the MoSi$_2$N$_4$ and WSi$_2$N$_4$ exhibit almost the same size of lattice thermal conductivity.

**IV CONCLUSIONS**

We investigate the lattice thermal conductivities of monolayer MoSi$_2$N$_4$ based on first-principles calculations and the Boltzmann transport equation. Unexpectedly, we find that its intrinsic lattice thermal conductivity (224 Wm$^{-1}$K$^{-1}$ at 300 K) are much higher than those of the other well-known semiconductors, such as black phosphorene, blue phosphorene, monolayer 2H-MoTe$_2$, and MoS$_2$. The detailed analysis indicates that the large lattice thermal conductivity of MoSi$_2$N$_4$ is attributed to the high phonon group velocities, long phonon lifetimes, and small Grüneisen parameters compared to the other well-known 2D materials. Besides, we compare the lattice thermal conductivities of monolayer $MoSi_2N_4$ and $WSi_2N_4$. It is found that the lattice thermal conductivity of $WSi_2N_4$ is only a little lower than those of monolayer $MoSi_2N_4$. We think that the high thermal conductivities of $MoSi_2N_4$ and $WSi_2N_4$ are possible due to their particular sandwiched structure, in which the Si-N layers are mainly responsible for the heat transportation. The high lattice thermal conductivities of monolayer $MoSi_2N_4$ and $WSi_2N_4$ making them promising building blocks for heat dissipation in nanoelectronics and microelectronics.

**ACKNOWLEDGMENTS**

This work is supported by the National Natural Science Foundation of China (Grant No. 11704195, 11974163, 11890702, and No. 51721001) and National Key R&D Program of China (Grant No. 2016YFA0201104). QFL also acknowledges the Qing Lan Project of Jiangsu Province (Grant No. R2019Q04). The numerical calculations in this paper have been done on the computing facilities in the High Performance Computing Center (HPCC) of Nanjing University.